\documentclass[letterpaper, 10 pt, conference]{ieeeconf}  

\IEEEoverridecommandlockouts                              
\usepackage[utf8]{inputenc}
\usepackage[T1]{fontenc}

\usepackage[bibstyle=ieee,
            citestyle=numeric-comp,
            minnames=1, maxnames=2,
            backend=biber,
            doi=false,isbn=false,
            url=true]{biblatex}
\AtEveryBibitem{\clearfield{month}}
\AtEveryCitekey{\clearfield{month}}

\newcommand{\etal}{\textit{et al}.}
\newcommand{\eg}{e.g.,}
\newcommand{\ie}{i.e.,}

\usepackage{graphicx}
\usepackage{nicefrac}
\usepackage{mathtools}
\usepackage{multicol, multirow, bigstrut}

\usepackage{amsmath} 
\usepackage{amssymb}  

\newcommand{\R}{\mathbb{R}}

\newcommand{\mat}[1]{\mathbf{#1}}
\renewcommand{\vec}[1]{\mathbf{\boldsymbol{#1}}}

\renewcommand{\cos}[1]{\mathbf{c}#1}
\renewcommand{\sin}[1]{\mathbf{s}#1}

\newcommand{\cs}[1]{#1}

\newcommand{\bvraww}[5]{\prescript{\cs{#1}}{#2}{\vec{#3}}^{#4}_{#5}}

\newcommand{\bvUR}[5]{
    \ifthenelse{\equal{#4}{} \OR \equal{#2}{}}{
        \bvraww{#1}{}{#3}{\cs{#2}#4}{#5}
    }{
        ( \bvraww{#1}{}{#3}{\cs{#2}}{#5} )^{#4}
    }
}
\newcommand{\bv}[5]{
    \ifthenelse{\equal{#1}{W}}{
        \bvUR{}{#2}{#3}{#4}{#5}
    }{
        \bvUR{#1}{#2}{#3}{#4}{#5}
    }
}

\newcommand\bovermat[2]{%
  \makebox[0pt][l]{$\smash{\overbrace{\phantom{%
    \begin{matrix}#2\end{matrix}}}^{\text{#1}}}$}#2}

\newcommand{\quatUR}[5]{
    \ifthenelse{\equal{#4}{} \OR \equal{#2}{}}{
        \prescript{\cs{#1}}{}{\boldsymbol{#3}}^{\cs{#2}#4}_{#5}
    }{
        (\prescript{\cs{#1}}{}{\boldsymbol{#3}}^{\cs{#2}}_{#5})^{#4}
    }
}
\newcommand{\quatURR}[5]{
    \ifthenelse{\equal{#1}{W}}{
        \quatUR{}{#2}{#3}{#4}{#5}
    }{
        \quatUR{#1}{#2}{#3}{#4}{#5}
    }
}

\usepackage{mdframed}
\usepackage{empheq}

\usepackage{xcolor, soul}
\iftrue
    \global\mdfdefinestyle{mdchange}{%
    backgroundcolor=yellow, linewidth=0pt,%
    leftmargin=0pt,rightmargin=0pt,
    skipabove=0,skipbelow=0,
    innerleftmargin=0pt,innerrightmargin=0pt,
    innertopmargin=0pt,innerbottommargin=0pt
    }

\else
    \global\mdfdefinestyle{mdchange}{%
    linewidth=0pt,%
    leftmargin=0pt,rightmargin=0pt,
    skipabove=0,skipbelow=0,
    innerleftmargin=0pt,innerrightmargin=0pt,
    innertopmargin=0pt,innerbottommargin=0pt
    }
    \sethlcolor{white}

\fi
\soulregister\cite7
\soulregister\ref7
\soulregister\pageref7
\soulregister\emph7

\title{\LARGE \bf
Estimating Lower Limb Kinematics using Distance Measurements with a Reduced Wearable Inertial Sensor Count
}

\author{Luke Sy$^1$ \emph{Student Member, IEEE}, 
        Nigel H. Lovell$^1$ \emph{Fellow, IEEE}, \\ 
        Stephen J. Redmond$^{1,2}$ \emph{Senior Member, IEEE} 
    \thanks{$^{1}$L. W. Sy, N. H. Lovell, and S. J. Redmond are with the Graduate School of Biomedical Engineering, UNSW Sydney, Australia \{\small l.sy, n.lovell, s.redmond\}@unsw.edu.au }%
    \thanks{$^{2}$S. J. Redmond is with the UCD School of Electrical and Electronic Engineering, University College Dublin, {\small stephen.redmond}@ucd.ie }%
}

\begin{document}
\maketitle
\thispagestyle{empty}
\pagestyle{empty}

\begin{abstract}
    This paper presents an algorithm that makes novel use of distance measurements alongside a constrained Kalman filter to accurately estimate pelvis, thigh, and shank kinematics for both legs during walking and other body movements using only three wearable inertial measurement units (IMUs).
        The distance measurement formulation also assumes hinge knee joint and constant body segment length, helping produce estimates that are near or in the constraint space for better estimator stability.
    Simulated experiments shown that inter-IMU distance measurement is indeed a promising new source of information to improve the pose estimation of inertial motion capture systems under a reduced sensor count configuration.
        Furthermore, experiments show that performance improved dramatically for dynamic movements even at high noise levels (\eg{} $\sigma_{dist} = 0.2$ m), 
        and that acceptable performance for normal walking was achieved at $\sigma_{dist} = 0.1$ m.
    Nevertheless, further validation is recommended using actual distance measurement sensors.
\end{abstract}

\section{Introduction}
	The study of human movement, \ie{} gait analysis, can help 
	    diagnose movement disorders (\eg{} falls risk \cite{Hausdorff2001}), assess the effect of surgery, 
	    and track the progression of a patient's rehabilitation \cite{Wren2011, Shull2014}.
    Human gait analysis involves the measurement of kinematic (\eg{} joint angles) and other related parameters (\eg{} stride length, joint moments) parameters.
        Traditionally, joint kinematics are captured within a laboratory setting using optical motion capture (OMC) systems which can estimate position with up to millimeter accuracy, if well-configured and calibrated.
	Recent miniaturization of inertial measurements units (IMUs), however, led to the development of inertial motion capture (IMC) systems.
	    IMCs can capture joint kinematics in unstructured environments and can be used to enable remote gait analysis.
	
	IMCs utilize either a one sensor per body segment (OSPS) or a reduced sensor count (RSC) configuration.
		Commercial IMCs use an OSPS configuration (\ie{} five IMUs to track the pelvis, thighs, and shanks) \cite{Roetenberg2009, LEGSys-BioSensics},
		    which may be considered too cumbersome and expensive for routine daily use by a consumer due to the number of IMUs required.
		    Each IMU typically tracks the orientation of the attached body segment using an orientation estimation algorithm (\eg{} \cite{DelRosario2016a, DelRosario2018}), which is then connected via linked kinematic chains, usually rooted at the pelvis.
		In a RSC configuration, IMUs are placed on a subset of body segments which can improve user comfort while also reducing setup time and cost.
			However, utilizing fewer IMUs inherently removes the kinematic information from the uninstrumented body segments which must be inferred from another source (\eg{} mechanical joint constraints, dynamic balance assumptions, or another kind of sensor).
		Acceptable measurement accuracies vary between clinical applications. 
		    Between the two configurations, OSPS typically have acceptable results ($<5^\circ$ RMSE bias removed and adjusted for repeatability) \cite{McGinley2009, Al-Amri2018},
		        while RSC performance needs improvement \cite{Sy2019-ckf}.

    RSC performance is dependent on how the algorithm infers the kinematic information of the body segments lacking attached sensors.
        One approach is to leverage our knowledge of human movement either through data obtained in the past (\ie{} observed correlations between co-movement of different body segments) or by using a simplified model of the human body.
        Data-driven approaches (\eg{} nearest-neighbor search \cite{Tautges2011} and bi-directional recurrent neural network \cite{Huang2018a}) are able to recreate realistic motion suitable for animation-related applications.
            However, these approaches are expected to have a bias toward motions already contained in the database, inherently limiting their use in monitoring pathological gait.
        On the other hand, model-based approaches reconstruct body motion using kinematic and biomechanical models
            (\eg{} inverse kinematics in 2D \cite{Hu2015}, constrained Kalman filter (KF) based \cite{Sy2019-ckf}, and window-based optimization \cite{Marcard2017}).
    
    Another approach is to supplement kinematic information from the IMU with another kind of sensor,
        which inherently increases cost and reduces battery life.
        Note that we will focus on systems that supplement pose estimate, not on the global position estimate of the subject (\eg{} \cite{Hol2009}).
            For example, IMCs can be supplemented with standard video cameras (\eg{} fused using an optimization-based algorithm \cite{Malleson2017}, and deep neural networks \cite{Gilbert2019}) or depth cameras \cite{Helten2013} at fixed locations in the capture environment, external to the subject.
                The combination of IMCs and portable cameras solves a weakness of OMCs, which is marker or body segment occlusion, and a weakness of IMCs, which is global position drift.
                However, the system still requires an external sensor that is carried by another person, or requires some quick setup.
            IMCs can also be supplemented by distance measurement (\eg{} using ultrasonic devices fused using an extended KF \cite{Vlasic2007}).
                Although to the best of the authors' knowledge, the use of distance measurements with IMCs only exist in an OSPS configuration.

    This paper builds on the authors' prior work \cite{Sy2019-ckf} and describes a novel algorithm based on a constrained KF (CKF) to estimate lower body kinematics using a RSC configuration of IMUs and inter-IMU distance measurements;
        the inclusion of inter-IMU distance measurements is the primary advancement made in this paper.
        This design was motivated by the need to develop a gait assessment tool using as few sensors as possible, ergonomically-placed for comfort, to facilitate long-term monitoring of lower body movement towards the tracking of all activities of daily living (ADL).

\section{Algorithm Description}
    This section will briefly describe our prior work which the new algorithm will be built on, followed by the proposed changes and formulation involved.
    
    \subsection{CKF-3IMU}
        The algorithm \emph{CKF-3IMU} is based on a CKF that estimates the orientation of the pelvis, thighs, and shanks with respect the world frame, $W$ \cite{Sy2019-ckf}. 
            At each time step, \emph{CKF-3IMU} predicts the position of the shanks and pelvis in 3D through double integration of their linear 3D acceleration as measured by the attached IMUs in a pre-processing step.
            To mitigate positional drift due to sensor noise that accumulates in the double integration of acceleration, the following assumptions are also enforced:
                (1) the ankle 3D velocity and height above the floor are zeroed whenever a footstep is detected;
                (2) the pelvis 3D position has an approximate Z position being the length of the unbent leg(s) above the floor, and an XY position being mid-way between the two ankle XY positions at any time.
            Lastly, there are biomechanical constraints that enforce constant body segment length; ball-and-sockets hip joints; and a hinge knee joint (one degree of freedom (DOF)) with limited range of motion (ROM).
            Note that \emph{CKF-3IMU} assumed that the orientation estimated from the IMUs in the pre-processing step is error-free. 
        
    \subsection{Proposed changes}
        Despite the strengths of \emph{CKF-3IMU}, the assumptions made to mitigate pelvis positional drift may not necessarily hold for movements other than walking.
            To infer kinematic information for the uninstrumented body segment that holds for most, if not all, ADL, while not utilizing any external sensor (\ie{} self contained on the person's body), 
            this paper proposes the novel use of sensors that measure distance between the existing IMUs attached on the pelvis and ankles.
            Such a sensor system could be implemented using ultrasonic transceivers \cite{Vlasic2007} or ultra-wideband (UWB) radio \cite{Hol2009}.
            
\begin{figure}[htbp]
            \centering
            \includegraphics[width=\linewidth]{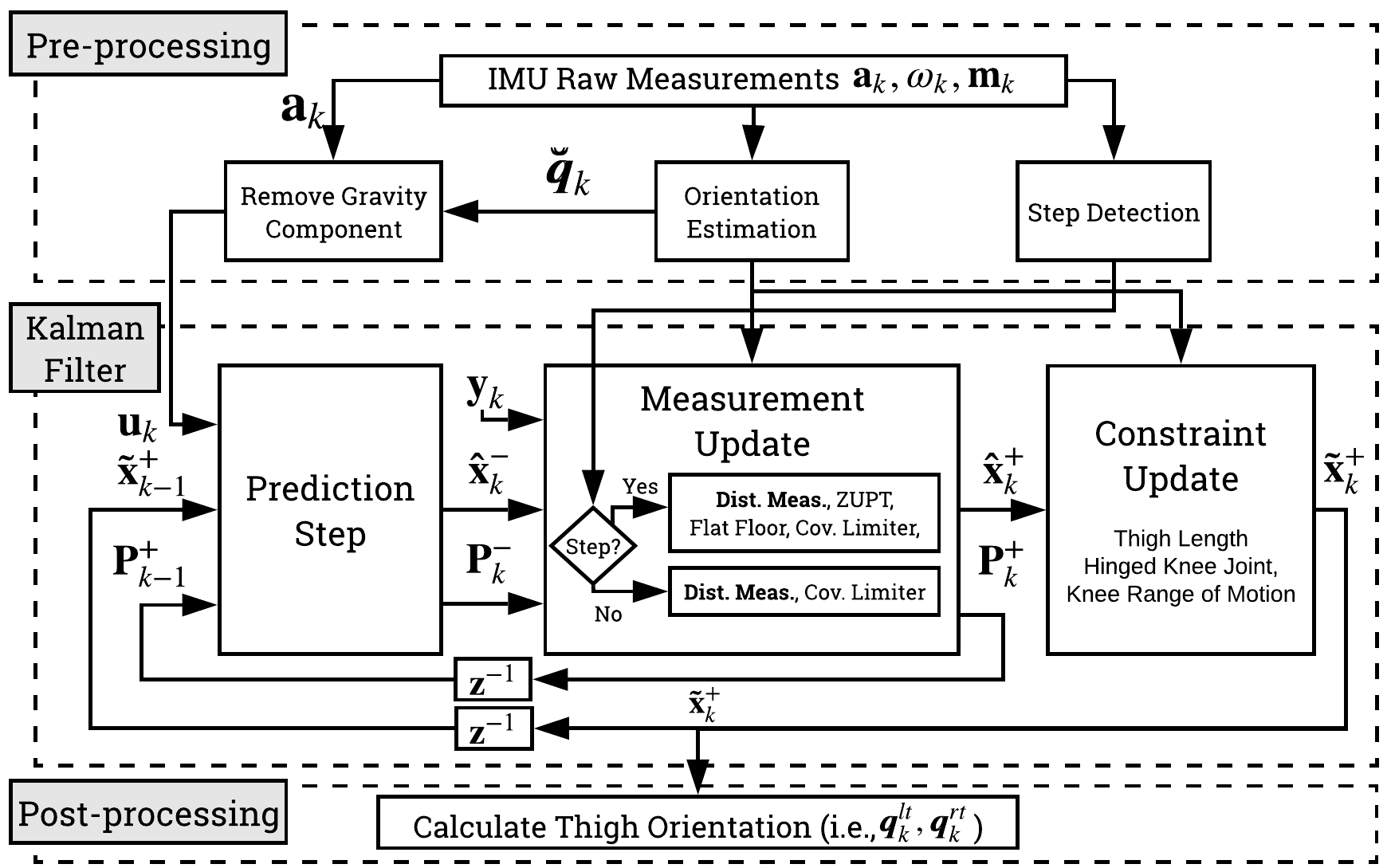}
            \caption{Algorithm overview which consists of three parts.
            (1) Pre-processing calculates the inertial acceleration, body segment orientation, and step detection from raw acceleration, $\vec{a}_{k}$, angular velocity, $\vec{\omega}_{k}$, and magnetic vector, $\vec{m}_{k}$, measured by the IMUs. 
            (2) The KF-based state estimation consists of a prediction (kinematic equation), measurement (distance measurements, covariance limiter, intermittent zero-velocity update (ZUPT), and flat-floor assumption), and constraint update (thigh length, hinge knee joint, and knee range of motion). 
            (3) Post-processing calculates the thigh orientations.}
            \label{fig:algo-overview}
        \end{figure}

        Using a similar physical body model and algorithm framework as \cite{Sy2019-ckf}, the additional distance measurements will be incorporated in the measurement update of the CKF (details in the next section) replacing the pelvis position pseudo measurements of \emph{CKF-3IMU}.
            Fig. \ref{fig:algo-overview} shows an overview of the proposed algorithm, denoted \emph{CKF-3IMU+D}. 
            Note that only the measurement update is changed. 
                The pre-processing, prediction step, constraint update, and post-processing remains exactly the same with \emph{CKF-3IMU} as described in Sections II-C, II-E.1, II-E.3, and II-F of \cite{Sy2019-ckf}, respectively.
        
    \subsection{Formulation}
\begin{figure}[htbp]
            \centering
            \includegraphics[width=0.35\textwidth]{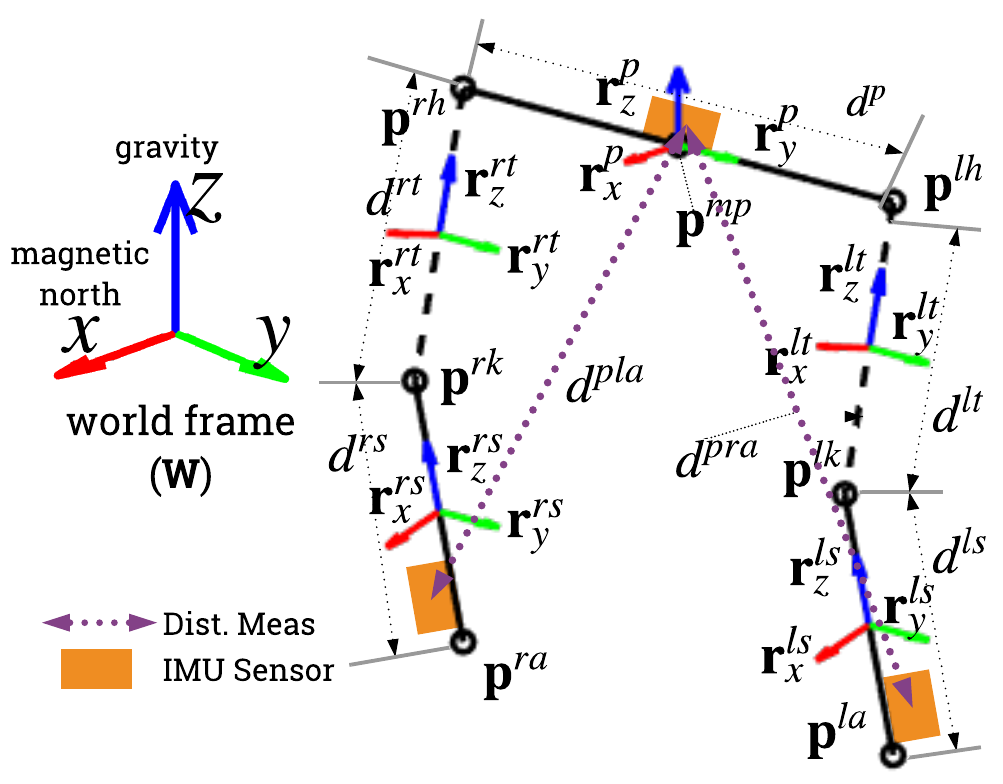}
            \caption{Physical model of the lower body used by the algorithm. The circles denote the joint positions.
            The solid lines denote instrumented body segments; 
            the dashed lines denote segments without IMUs attached (\ie{} the thighs);
            and the dotted lines denote distance measurements. 
            Estimating the orientation and position of the uninstrumented thighs is the main objective of this paper.}
            \label{fig:body-skeleton-sparse-ckf}
        \end{figure}

        The model used for the CKF can be defined as follows:
    	    \begin{gather}
    	        \vec{x}_{k} = \mat{F} \vec{x}_{k-1} + \mat{G} \vec{u}_{k-1} + \vec{w}_{k-1} \label{eq:pred-update} \\
    	        \vec{y}_{k} = \mat{H}_{k} \vec{x}_{k} + \vec{v}_{k}, \qquad 
    	        \mat{D}_{k} \vec{x}_{k} = \vec{d}_{k} \label{eq:meas-const-update}
            \end{gather}
        where $k$ is the time step; 
            $\vec{x}_{k}$ and $\vec{y}_{k}$ are the state and measurement vector, respectively; 
            $\vec{w}_{k}$ and $\vec{v}_{k}$ are zero-mean process and measurement noise vectors, respectively, with covariance matrices $\mat{Q}_{k}$ and $\mat{R}_{k}$;
            $\mat{F}$, $\mat{G}$, and $\mat{H}_{k}$ are the state transition, input, and measurement matrices, respectively; 
            and $\mat{D}_{k} \vec{x}_{k} = \vec{d}_{k}$ are the equality constraints that state $\vec{x}_{k}$ must satisfy.
        The state variables in $\vec{x}_{k}$ model the position and velocity of the instrumented body segments (\ie{} the $18 \times 1$ vector $\vec{x}_{k}$ = $\big[\bv{W}{mp}{p}{T}{k}$, $\bv{W}{la}{p}{T}{k}$, $\bv{W}{ra}{p}{T}{k}$, $\bv{W}{mp}{v}{T}{k}$, $\bv{W}{la}{v}{T}{k}$, $\bv{W}{ra}{v}{T}{k} \big]^{T}$).
            $\bv{D}{C}{p}{}{}$ and $\bv{D}{C}{v}{}{} \in \R^3$ denotes the 3D position and velocity of joint $\cs{C}$ with respect to frame $\cs{D}$. 
                If frame $\cs{D}$ is not specified, assume reference to the world frame, $\cs{W}$.
            See Fig. \ref{fig:body-skeleton-sparse-ckf} for visualization.
            Note that $\bv{W}{S}{r}{}{x}$, $\bv{W}{S}{r}{}{y}$, and $\bv{W}{S}{r}{}{z} \in \R^3$ are the basis vectors of the segment's local frame, $\cs{S}$, with respect to the world frame, $\cs{W}$.

        The proposed changes to the algorithm will primarily affect the measurement matrix and vector, $\mat{H}_{k}$ and $\vec{y}_{k}$ (see next section).
            Refer to \cite{Sy2019-ckf} for the full formulation of the original algorithm.
        The first step of \mbox{\emph{CKF-3IMU+D}} will be the prediction step which implements standard kinematic equations (similar formulation to \emph{CKF-3IMU}) and is followed by the measurement update.
        
    \subsubsection{Measurement update} \label{sec:meas-update}
        estimates the next state by: 
            (i) infering pelvis position from distance measurements while assuming hinged knee joints and constant body segment lengths (\ie{} state variables are in the constraint subspace), and by; 
            (ii) utilizing zero ankle velocity and flat floor assumptions whenever a footstep is detected.
            For the covariance update, there is an additional step to limit the \textit{a posteriori} error covariance matrix elements from growing indefinitely and from becoming badly conditioned.
        The distance measurement formulation will be described in this subsection.
        Refer to \cite[Sec. II-E.2]{Sy2019-ckf} for the formulation of the zero ankle velocity, flat floor assumption, and covariance limiter.

        Assuming a hinged knee joint and constant body segment length, the pelvis to left ankle vector can be calculated from the pelvis to left ankle distance,
            and incorporated into our measurement model via $\mat{H}_{pla}$ and $\vec{y}_{pla}$, as shown in Eq. \eqref{eq:H-y-distl}, with measurement noise variance $\bv{}{}{\sigma}{2}{pla}$ ($3 \times 1$ vector).
            For the sake of brevity, only the left side formulation is shown. 
            The right side (\ie{} pelvis to right ankle vector) can be calculated similarly.
            Note that the measurement model could also be formulated as the linearized Euclidean distance between the pelvis and ankle. 
                However, preliminary exploration of this approach showed poor performance.
                More details can be found in the discussion. 
        
        First, we solve for an estimated left knee angle, $\hat{\theta}^{+}_{lk}$, from the measured pelvis to left ankle distance, $\hat{d}^{pla}$.
            The pelvis to left ankle vector, $\bv{W}{pla}{\tau}{}{k}(\theta_{lk})$ (Eq. \eqref{eq:dist-pla-vec}), can be defined as the sum of the mid-pelvis to hip, thigh, and shank vectors.
            Note that $\cos{}$ and $\sin{}$ denote \textbf{cos} and \textbf{sin},
                $\cdot$ denotes the dot product operator,
                and $\vec{a}^2$ denotes $\vec{a} \cdot \vec{a}$.
            By definition of $(d^{pla})^2$ and expanding $\bv{W}{pla}{\tau}{}{k}(\theta_{lk})$ with Eq. \eqref{eq:dist-pla-vec}, Eq. \eqref{eq:dist-pla-mag-base} is obtained and can be rearranged in the form of Eq. \eqref{eq:dist-pla-form} with $\alpha$, $\beta$, $\gamma$ as shown in Eqs. \eqref{eq:dist-pla-form-ab} and \eqref{eq:dist-pla-form-c}.
            Solving for $\hat{\theta}_{lk}$ from Eq. \eqref{eq:dist-pla-form} gives us a quadratic equation with two solutions as shown in Eq. \eqref{eq:dist-pla-thetalk}.
            Between the two solutions, $\hat{\theta}^{+}_{lk}$ is set as the $\hat{\theta}_{lk}$ whose value is closer to the current left knee angle estimate from the CKF prediction step.
            Thus, given an assumed perfect measurement of pelvis and shank orientation, and assuming a hinged knee joint and ball-and-socket hip joint, we have found the knee angle which gives a pelvis to left ankle distance which best matches the distance measured by our new distance sensor; this solution serves as a pseudomeasurement of the knee angle.
        Finally, $\vec{y}_{pla}$, the KF measurement, is the inter-IMU vector between the pelvis and left ankle, calculated using Eq. \eqref{eq:dist-pla-vec} with input $\hat{\theta}_{lk}^{+}$.
        \begin{gather}
            \vec{\psi} = \tfrac{d^{\cs{p}}}{2} \bv{W}{p}{r}{}{y} - d^{\cs{ls}} \bv{W}{ls}{r}{}{z} \\
            \bv{W}{pla}{\tau}{}{k}(\theta_{lk}) = 
                \overbrace{\vec{\psi}}^{\text{hip+shank}}
                + \overbrace{d^{\cs{lt}}\bv{W}{ls}{r}{}{x} \sin{(\theta_{lk})}
                            -d^{\cs{lt}}\bv{W}{ls}{r}{}{z} \cos{(\theta_{lk})}
                            }^{\text{thigh}} 
                            \label{eq:dist-pla-vec}
        \end{gather}
        
        \begin{gather}
            (\hat{d}^{pla})^2 = \bv{W}{pla}{\tau}{}{k}(\theta_{lk})^2 
                = \vec{\psi}^2 - 2 d^{\cs{lt}} \vec{\psi} \cdot \bv{W}{ls}{r}{}{z} \cos{(\theta_{lk})} \nonumber \\ 
                    + 2 d^{\cs{lt}} \vec{\psi} \cdot \bv{W}{ls}{r}{}{x} \sin{(\theta_{lk})} + (d^{lt})^2 \label{eq:dist-pla-mag-base} \\
            \alpha \cos{(\theta_{lk})} + \beta \sin{(\theta_{lk})} = \gamma \label{eq:dist-pla-form} \\
            \alpha = - 2 d^{\cs{lt}} \vec{\psi} \cdot \bv{W}{ls}{r}{}{z}, \quad
            \beta = 2 d^{\cs{lt}} \vec{\psi} \cdot \bv{W}{ls}{r}{}{x} \label{eq:dist-pla-form-ab}\\
            \gamma = (\hat{d}^{pla})^2 - \vec{\psi}^2 - (d^{lt})^2 \label{eq:dist-pla-form-c} \\
            \hat{\theta}_{lk} = \mathbf{c}^{-1}\left( \tfrac{\alpha \gamma \pm \beta \sqrt{\alpha^2+\beta^2-\gamma^2}}{\alpha^2+\beta^2} \right) \label{eq:dist-pla-thetalk} \\[1em]
            \mat{H}_{pla} = \left[ \begin{array}{rrr}
                \bovermat{pelv. pos.}{ -\mat{I}_{3 \times 3} } & \bovermat{\emph{la} pos.}{ \mat{I}_{3 \times 3} } & ... \\
                \end{array} \right], \:\:
            \vec{y}_{dl,k} = \bv{W}{pla}{\tau}{}{k}(\hat{\theta}^{+}_{lk}) \label{eq:H-y-distl}
        \end{gather}
        Lastly, the modified measurement matrix $\mat{H}_{k}$ varies with time through a case statement depending on floor contact (FC)  with the foot as shown in Eq. \eqref{eq:H-k-cases}.
            Measurements $\vec{y}_{k}$ and measurement variances  $\bv{}{}{\sigma}{2}{k}$ are constructed similarly.
            \begin{align}
                \mat{H}_{k} = \begin{cases}
                    [ \mat{H}_{pla}^T \quad \mat{H}_{pra}^T ]^T & \text{ no foot FC} \\
                    [ \mat{H}_{pla}^T \quad \mat{H}_{pra}^T \quad \mat{H}_{ls}^T ]^T & \text{ left foot FC} \\
                    [ \mat{H}_{pla}^T \quad \mat{H}_{pra}^T \quad \mat{H}_{rs}^T ]^T & \text{ right foot FC} \\
                    [ \mat{H}_{pla}^T \quad \mat{H}_{pra}^T \quad \mat{H}_{ls}^T \quad \mat{H}_{rs}^T ]^T & \text{ both feet FC} \\
                \end{cases} \label{eq:H-k-cases}
            \end{align}

        After the measurement update is the constraint update which enforces biomechanical constraints with similar formulation to \emph{CKF-3IMU}.
        
\section{Experiment}
        The dataset from \cite{Sy2019-ckf} was used to evaluate both algorithms.
        The evaluation involved movements listed in Table \ref{tab:movement-type-desc} from nine healthy subjects ($7$ men and $2$ women, weight $63.0 \pm 6.8$ kg, height $1.70 \pm 0.06$ m, age $24.6 \pm 3.9$ years old), with no known gait abnormalities.
            Raw data were captured using a commercial IMC (\ie{} Xsens Awinda) compared against a benchmark OMC (\ie{} Vicon) within an approximately $4 \times 4$ m$^2$ capture area.
            Unless stated, calibration and system parameters similar to \cite{Sy2019-ckf} were assumed.
        \begin{table}[htbp]
            \centering
            \caption{Types of movements done in the validation experiment}
            \label{tab:movement-type-desc}
            \begin{tabular}{lllc}
                \hline \hline
                 Movement & Description & Duration & Group \\ \hline 
                 Walk & Walk straight and return & $\sim 30$ s & F\\
                 Figure-of-eight & Walk along figure-of-eight path & $\sim 60$ s & F \\
                 Zig-zag & Walk along zig-zag path & $\sim 60$ s & F \\
                 5-minute walk & Unscripted walk and stand & $\sim 300$ s & F \\
                 TUG & Timed up and go & $\sim 30$ s & D \\
                 Speedskater & Speedskater on the spot & $\sim 30$ s & D \\
                 Jog & Jog straight and return & $\sim 30$ s & D \\
                 Jumping jacks & Jumping jacks on the spot & $\sim 30$ s & D \\
                 High knee & High knee jog on the spot & $\sim 30$ s & D \\
                 \hline \hline
            \end{tabular}
            F denotes free walk, D denotes dynamic
        \end{table}
    
        The distance measurements, $d^{\cs{pla}}$ and $d^{\cs{pra}}$, were simulated by calculating the distance from the mid-pelvis to the left and right ankles and adding normally distributed positional noise with different standard deviations 
            (\ie{} $\sigma_{dist} = [0,0.01,\dots,0.1,0.15,0.2]$ m).
    
        Lastly, the evaluation was done using the following metrics:
        (1) joint angles RMSE with bias removed and coefficient of correlation (CC) of the hip in the sagittal, frontal, and transverse planes and of the knee in the sagittal plane;
        and (2) Total travelled distance (TTD) deviation (\ie{} TTD error with respect to the actual TTD) of the ankles.
        Refer to \cite[Sec. III]{Sy2019-ckf} for more details.
            
\section{Results}
\begin{figure}[htbp]
        \centering
        \includegraphics[width=\linewidth]{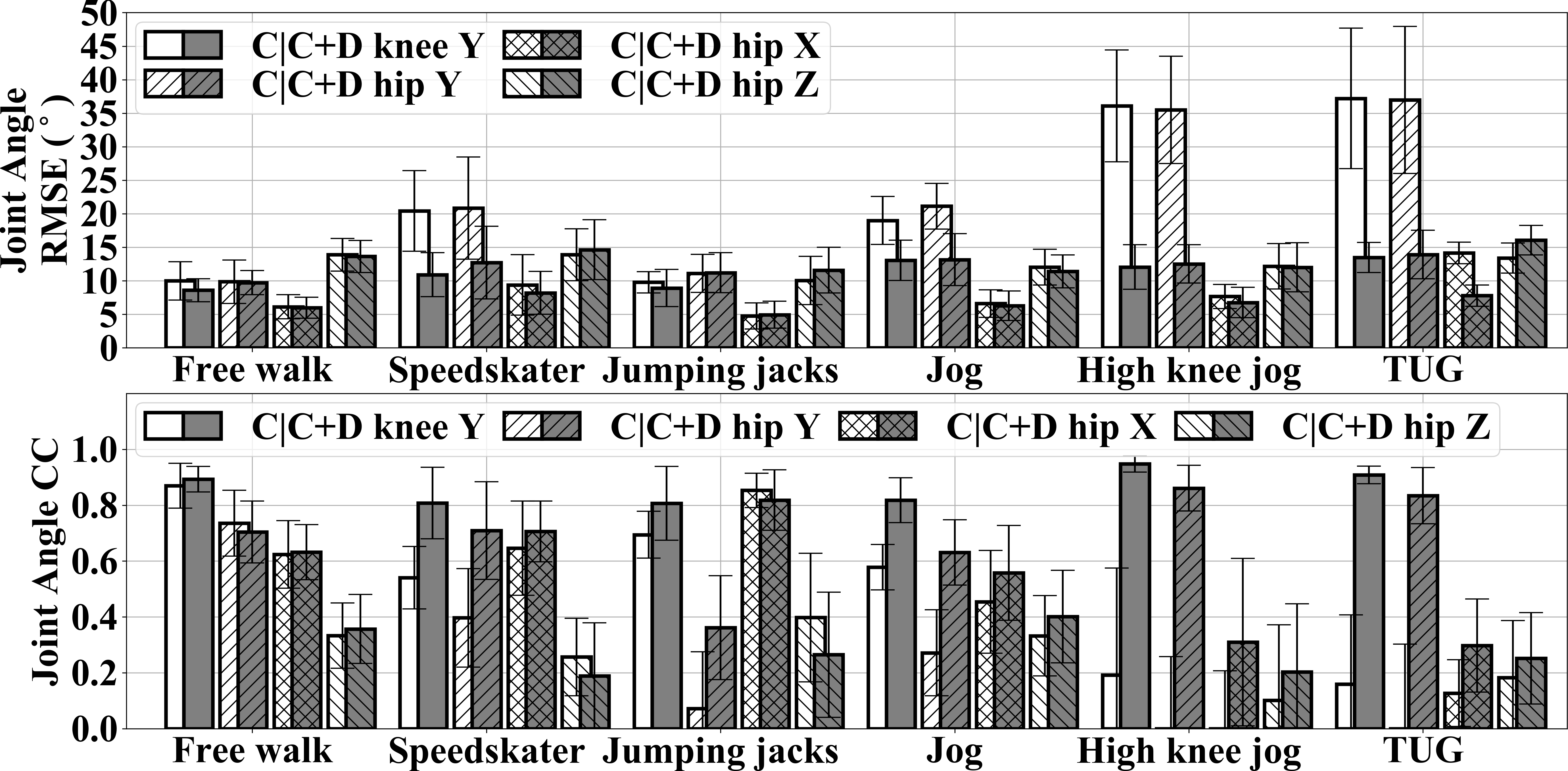}
        \caption{The joint angle RMSE (top) and CC (bottom) of the knee and hip at each motion type. Y, X, and Z denotes the sagittal, frontal, and transversal plane, respectively.
        The prefix C denotes \emph{CKF-3IMU} algorithm, while C+D denotes \emph{CKF-3IMU+D} algorithm.
        Note that in general, the gray boxes (\emph{CKF-3IMU+D} results) had lower RMSE and higher CC than the white boxes (\emph{CKF-3IMU} results).}
        \label{fig:results-kneehip-angles-rmsecc}
    \end{figure}

\begin{figure}[htbp]
        \centering
        \includegraphics[width=\linewidth]{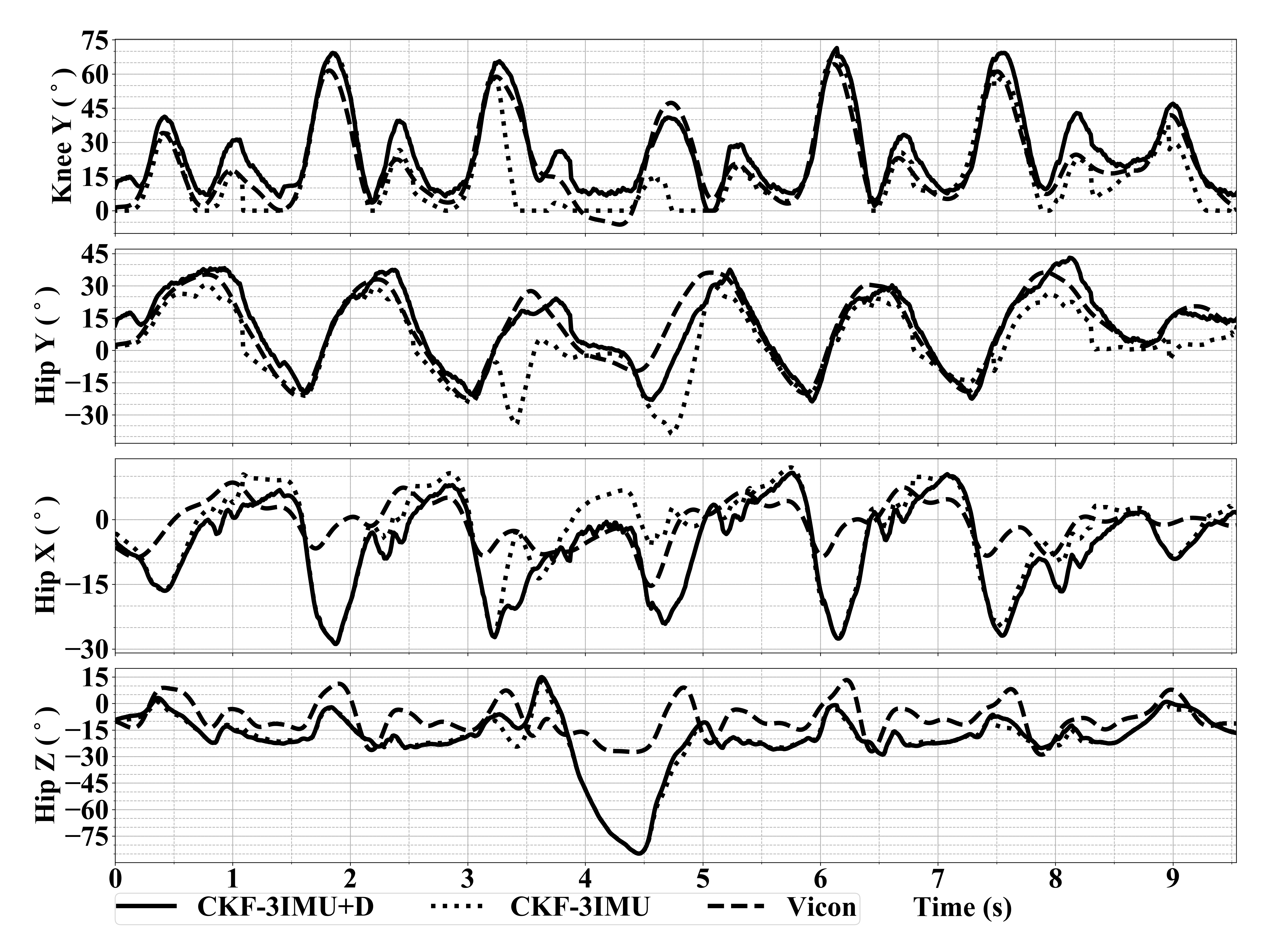}
        \caption{Knee and hip joint angle output of \emph{CKF-3IMU+D} and \emph{CKF-3IMU} in comparison to a benchmark OMC (Vicon) for a \emph{Walk} trial. The subject walked straight from $t=0$ to $3$ s, turned $180^\circ$ around from $t=3$ to $5.5$ s, and walked straight to original point from $5.5$ s until the end of the trial.}
        \label{fig:results-kneehip-angle-sample}
    \end{figure}

\begin{figure}[htbp]
        \centering
        \includegraphics[width=\linewidth]{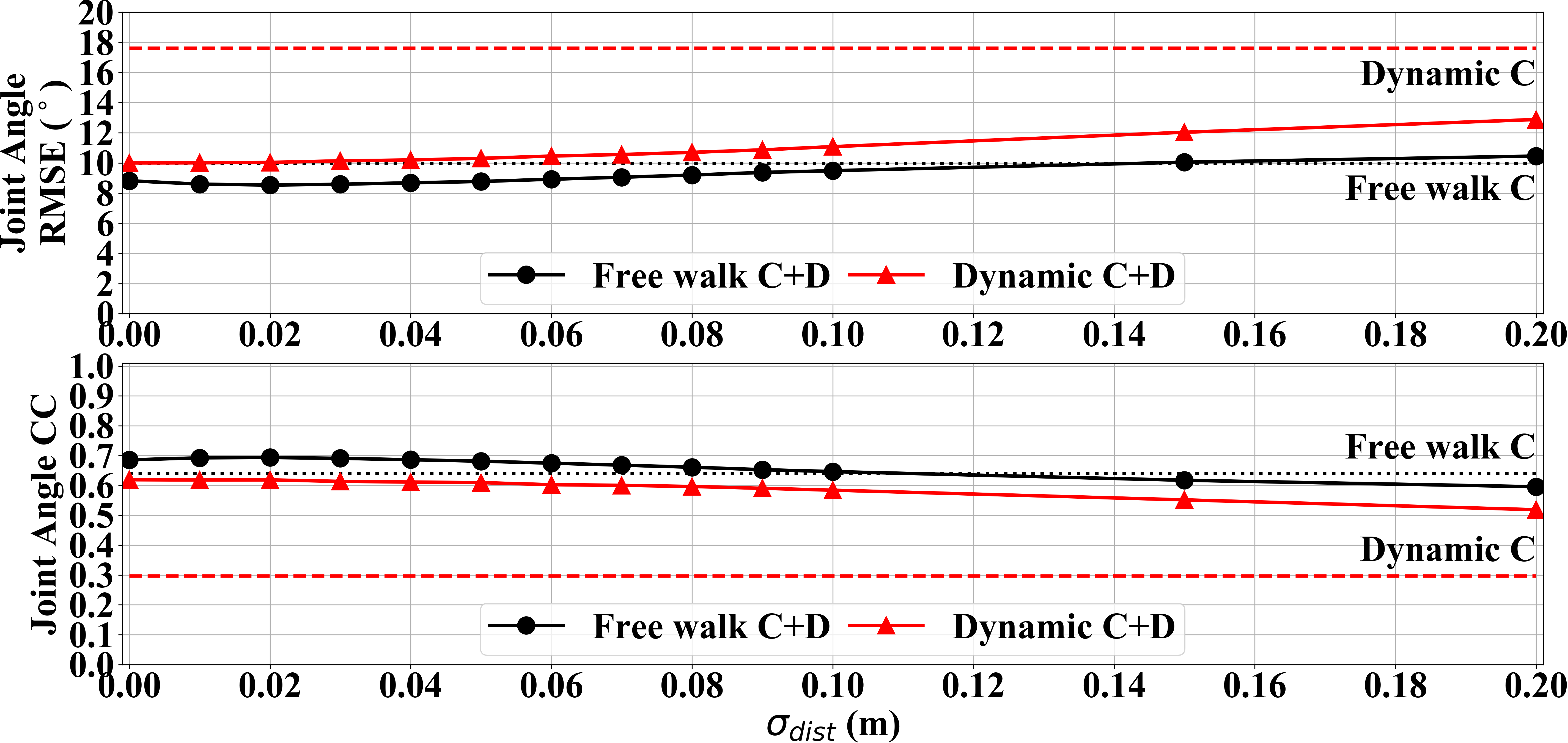}
        \caption{The joint angle RMSE (top) and CC (bottom) of free walk and dynamic movements at different noise level $\sigma_{dist}$.
        The suffix C denotes \emph{CKF-3IMU}, while C+D denotes \emph{CKF-3IMU+D}.}
        \label{fig:results-varysigma}
    \end{figure}

    Fig. \ref{fig:results-kneehip-angles-rmsecc} shows the knee and hip joint angle RMSE and CC at $\sigma_{dist}=0.1$ m.
        Y, X, and Z refers to the sagittal, frontal, and transversal planes, respectively.
    Fig. \ref{fig:results-kneehip-angle-sample} shows a sample \emph{Walk} trial.
    Table \ref{tab:results-ttddev} shows the TTD deviation at the ankles for selected movements that had significant XY displacement (\eg{} jumping jacks on the spot was not included).
    Fig. \ref{fig:results-varysigma} shows the total (mean of knee and hip) joint angle RMSE and CC at different $\sigma_{dist}$ values.
    Refer to the supplementary material for a video reconstructions of sample trials \cite{Sy-ckfdist2020-supp}.
    
    \begin{table}[htbp]
        \centering
        \caption{Total travelled distance (TTD) deviation from optical motion capture (OMC) system at the ankles}
        \begin{tabular}{lrrrr}
        \hline
        \hline
              & \multicolumn{2}{c}{\emph{CKF-3IMU}} & \multicolumn{2}{c}{\emph{CKF-3IMU+D}} \bigstrut[t]\\
              & \multicolumn{1}{c}{Left} & \multicolumn{1}{c}{Right} & \multicolumn{1}{c}{Left} & \multicolumn{1}{c}{Right} \bigstrut[b]\\
        \hline
        Free walk & 3.81\% & 3.60\% & 1.90\% & 1.85\% \bigstrut[t]\\
        Jog   & 24.02\% & 28.14\% & 26.66\% & 30.17\% \\
        TUG   & 6.17\% & 7.82\% & 3.18\% & 3.53\% \bigstrut[b]\\
        \hline
        \hline
        \end{tabular}%
        \label{tab:results-ttddev}%
    \end{table}%

\section{Discussion}
    Fig. \ref{fig:results-kneehip-angles-rmsecc} shows that although there was minimal hip and knee joint angle RMSE and CC improvement for free walk between \emph{CKF-3IMU} and \emph{CKF-3IMU+D}, there was significant improvement for most dynamic movements, specifically, speedskater, jog, high knee jog, and TUG.
        Further observation shows that the CC for dynamic movements started to reach similar performance with free walk movement, indicating that distance measurements has indeed made the pose estimator capable to track more ADLs and not just walking.
        The hip and knee joint angle RMSE and CC were also comparable to the performance of OSPS based systems \cite{Sy2019-ckf, Cloete2008}.
            Similar to IMC based systems, \emph{CKF-3IMU+D} also follows the trend of having sagital (Y axis) joint angles similar to that capture by OMC systems, but with significant difference in frontal and transverse (X and Z axis) joint angles \cite{Cloete2008}.
        The hip Y and knee Y joint angle RMSE and CC were also slightly worse than Hu \mbox{\etal{}}.
            However, their validation was limited to a single gait cycle of walking, in contrast to the longer-duration walking trials and dynamic motions in this paper.
    The improvements can also be observed in the sample trial (Fig. \ref{fig:results-kneehip-angle-sample}), specifically from $t=3$ to $5$ s where Knee Y angle straightens out (\ie{} becomes $0^\circ$) early for \emph{CKF-3IMU} but follows a more similar pattern with OMC for \emph{CKF-3IMU+D},
        indicating that the distance measurements helped during the turning motion where the pelvis assumptions of \emph{CKF-3IMU} probably did not hold.

    Although distance measurements only provide relative distance between sensors, Table \ref{tab:results-ttddev} shows that \emph{CKF-3IMU+D} improved TTD for free walk and TUG.
    
    Fig. \ref{fig:results-varysigma} shows that as the simulated distance measurement noise $\sigma_{dist}$ becomes smaller, both joint angle RMSE and CC expectedly improve, albeit gradually, for both free walk and dynamic movements.
        Interestingly, distance measurement still dramatically improves performance even at higher noise levels (\eg{}  $\sigma_{dist} = 0.2$ m) for dynamic movements.
        However, the benefit plateaus at around $\sigma_{dist} = 0.05$ m.
        It was found experimentally that $\sigma_{dist} = 0.1$ m produced similar performance with \emph{CKF-3IMU} for free walk movements,
            which suggests that the distance measurement sensor noise must be $\sigma_{dist} \leq 0.1$ to reach performance that is not worse than the pelvis position pseudo-measurements utilized by \emph{CKF-3IMU}.
    We also attempted using raw distance measurements as a soft constraint in the measurement update.
        However, the resulting performance was worse that \emph{CKF-3IMU} even at negligible noise levels ($\sigma_{dist}$), probably because of the contention it brings with the constraint update where such raw distance measurements move the state outside the constraint subspace, only to be returned to the constraint subspace during the constraint update.
    
    Despite the promising performance when using distance measurements, further validation is needed in an actual hardware implementation, as the noise in the real world may not necessarily follow a normal distribution and may be non-stationary.
        For reference, portable ultrasound-based distance measurement can achieve accuracy in the range of millimetres with a sampling rate of $125$ Hz \cite{Vlasic2007}, 
            while a commercial UWB-based distance measurement device can achieve an accuracy in the range of $10$ cm with a sampling rate of $200$ Hz \cite{Malajner2015, Ledergerber2017}.
        Furthermore, combining distance measurements with a pose estimator that also tracks orientation may better utilize the new distance information leading to improved performance.

\section{Conclusion}
    Distance measurement is a promising new source of information to improve the pose estimation of IMC under RSC configuration.
        Simulations show that performance improved dramatically for dynamic movements even at lower noise levels (\eg{} $\sigma_{dist} = 0.2$m), 
        and that similar performance to \emph{CKF-3IMU} was achieved at $\sigma_{dist} = 0.1$ m for free walk movements.
    However, further validation is recommended with actual distance measurement sensors.
    The source code for the \emph{CKF-3IMU+D} algorithm and a sample video will be made available at \url{https://git.io/JvLCF}.





\section*{ACKNOWLEDGMENT}
This research was supported by an Australian Government Research Training Program (RTP) Scholarship.

\section*{References}
\printbibliography[heading=none]

\end{document}